\documentclass[12pt]{iopart}

\usepackage{iopams}  
\usepackage[dvips]{graphicx}
\usepackage{epsfig}
\usepackage{psfrag}

\begin{document}

\title{Clock synchronization by remote detection of correlated photon pairs}

\author{Caleb Ho$^{1}$, Ant\'{\i}a Lamas-Linares$^{1,2}$ and Christian Kurtsiefer$^{1,2}$}

\address{$^1$ Centre for Quantum Technologies, National University of Singapore,
3 Science Drive 2, Singapore, 117543}
\address{$^2$ Department of Physics, National University of Singapore,
2 Science Drive 3, Singapore, 117542}

\ead{christian.kurtsiefer@gmail.com}

\date{\today}

\begin{abstract}
We present an algorithm to detect the time and frequency difference of
independent clocks based on observation of time-correlated photon pairs.
This enables remote coincidence identification in entanglement-based
quantum key distribution schemes without dedicated coincidence hardware,
pulsed sources with a timing structure or very stable reference clocks.
We discuss the method for typical operating conditions, and show
that the requirement in reference clock accuracy can be
relaxed by about 5 orders of magnitude in comparison with previous schemes.
\end{abstract} 

\pacs{
03.67.Hk, 
07.05.Kf,  
95.99.Sh, 
42.65.Lm  
}

\maketitle

\section{ Introduction}
Quantum key distribution (QKD) \cite{scarani:08, gisin:07, dusek:06} is the
only quantum information protocol that found its way into
practical applications, and is currently in a stage of early commercial
development. There are two families of protocols that use fundamentally
different resources. The original QKD protocol BB84 \cite{bennett:84} and its
variants transmit single photons (or approximations thereof), while the other
family \cite{ekert:91} perform measurements on pairs of entangled photons. 
A few years ago, entanglement-based QKD protocols were viewed as
equivalent to BB84 \cite{bennett:92},
and thus only of little interest for practical QKD due to their
additional complexity. The new concept of device-independent QKD
\cite{acin:07}, and a returned awareness of classical side channels in
prepare-and-send protocols
revived interest in entanglement-based QKD schemes.
Entangled photon pairs are efficiently prepared by spontaneous parametric down
conversion (SPDC). Demonstrated for polarization-entangled pairs in 1995
\cite{kwiat:95}, recent developments lead to the extremely bright sources
available today \cite{fedrizzi:07, trojek:08}, so that entanglement-based QKD
became a viable option. 

The first step in establishing a key in such a scheme is the assignment of
photodetection events to entangled photon pairs. Due to their 
strong temporal correlation (down to a few 100\,fs) in typical
pair sources \cite{burnham:70}, this assignment can be done via
temporal coincidence identification. In typical laboratory experiments, as well as in 
early QKD implementations, a hardware channel was used to
carry out this coincidence identification \cite{poppe:04, peng:05}.
Less hardware is required when coincidences are identified
by comparing detection times given by good local clocks
\cite{jennewein:00,marcikic:06} or a 
central GPS time reference \cite{ursin:07}.

In this paper, we present an algorithm that relaxes the rather stringent
reference clock quality requirements for such a coincidence identification so
that  
conventional crystal oscillators can be used. In section~\ref{sec:tracking},
we outline the general problem and present a robust coincidence tracking
scheme. 
Section~\ref{sec:simpletimediff} covers the algorithm to find an
initial time offset as implemented in earlier experiments \cite{marcikic:06,
  ling:08}. In sections \ref{sec:twodiffs}-\ref{sec:finepart} we extend this
scheme in the presence of a frequency difference between the clocks necessary
to permit the use of clocks with lower accuracy.

\section{Photon pair identification with remote clocks}
\label{sec:tracking}
\begin{figure}
  \centerline{ \includegraphics{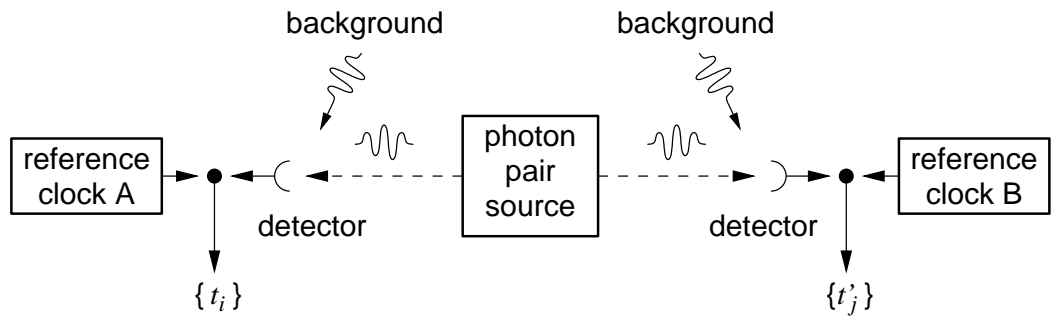}}
\caption{\label{fig:setting}
  Setting of the problem. Detection times of photoevents from a correlated
  photon pair source and background are registered with respect to two local
  reference clocks at remote locations A, B. The true coincidences 
  need then to be identified from
  the time sets $\{t_i\}$ and $\{t'_j\}$ on both sides.
}
\end{figure}

The identification of pairs is straightforward in any context in which a
hardware coincidence gate can be used; this is the case in laboratory-based
experiments or field setups with a dedicated synchronization channel. 

The situation we address in this paper applies to cases where detection times
of photons at the two distant locations \cite{marcikic:06, ursin:07, erven:08}
are recorded, and coincidences are identified based on these time stamps (see
figure~\ref{fig:setting}). This method
requires stable and synchronous clocks used for the timestamping: A
typical coincidence window $\tau_c$ is chosen to be slightly larger than the
detector 
time jitter, which is on the order of 1\,ns. The data acquisition for
establishing a key out of measurements is supposed to run either continuously,
or at least for a few 100 seconds. To maintain two clocks synchronized within
$\tau_c$ after a time of 100\,s, a relative accuracy of $10^{-11}$ is
required, a specification that is met by  commercial Rubidium clocks. For
longer operation times, this still may be insufficient unless either a timing
signal is transmitted on a separate channel, or the time reference is provided
by a central source. 

Pair sources based on SPDC provide enough information
in the streams of photodetection times $\{t_i\}$ and $\{t'_j\}$ that
such accurate clocks should not be necessary. As long as the pair events are
initially identified, the drift of the clocks can be tracked directly
from the coincidence signal. For this to work reliably, the rate of pair
events must be significantly larger than the one for accidental coincidences
due to background photons in the same time window $\tau_c$, which is also a
necessary condition for a obtaining a secure key in QKD.
.

In its simplest form, a floating average of the time difference $\Delta t=t_i-t'_j$
between true coincidence events can be used to track a drift of the reference
time between the two sides. To illustrate this, and to evaluate the intrinsic clock stability necessary to
follow the coincidence signature, we consider a realistic situation where
the full width at half maximum of a coincidence time distribution due to
detector jitter is $\tau_d=1$\,ns. To estimate the center of this distribution
with an uncertainty (one standard deviation) of $\delta\tau=0.1$\,ns, we
need to average time differences over about
\begin{equation}
n=\left({\tau_d\over2\sqrt{2\ln2}\delta\tau} \right)^2+1\approx 19
\end{equation}
coincidence events. Even for very low coincidence detection rates of
100 counts per second (cps), it takes less than 0.2 seconds to get a
sufficient number of 
events. Over that period, the clock should not drift such that an event leaves the
coincidence window, which translates into a relative frequency {\em accuracy}
requirement of $10^{-8}$ over 100\,ms. More realistic coincidence
detection rates of 1-10\,kcps require only a relative frequency accuracy of
$10^{-7}$ to $10^{-6}$ over a period of 1 to 10\,ms. Standard crystal
oscillators easily exhibit a {\em stability} on that order, but may lack the
accuracy. Thus, tracking the time difference in coincidences from a set of
detection events permits to use these simpler reference oscillators 
during normal operation.

Two problems are left for recovering the coincidences from
time stamps derived with respect to two separate clocks: First, the detection
instances  at both sides will have an unknown time offset 
$\Delta T$ between them. This is mainly due to the absence of a common
origin of time with a high enough resolution, and propagation over the physical
distance between the two sides. 
As long as two reference clocks have the same frequency, $\Delta T$ can be
found by looking at the cross correlation between the two timing signals. We
will elaborate this in the next section.

The second problem is related to the relative frequency difference between the
two clocks due to a lack of accuracy. This is harder to solve, since the
stream of time stamps $\{t_i\}$ and $\{t'_j\}$ on each side has no
intrinsic time structure: Both signal and background events follow a Poisson
distribution\footnote{While it has been shown that 
  the light emerging from SPDC processes should exhibit super-poissonian
  statistics \cite{holmes:89}, this is typically not observed in practical SPDC
  systems with photon counting detectors because it is washed out by
  multi-mode effects on a time scale way below the detector resolution.}.

\begin{figure} \centerline{\epsffile{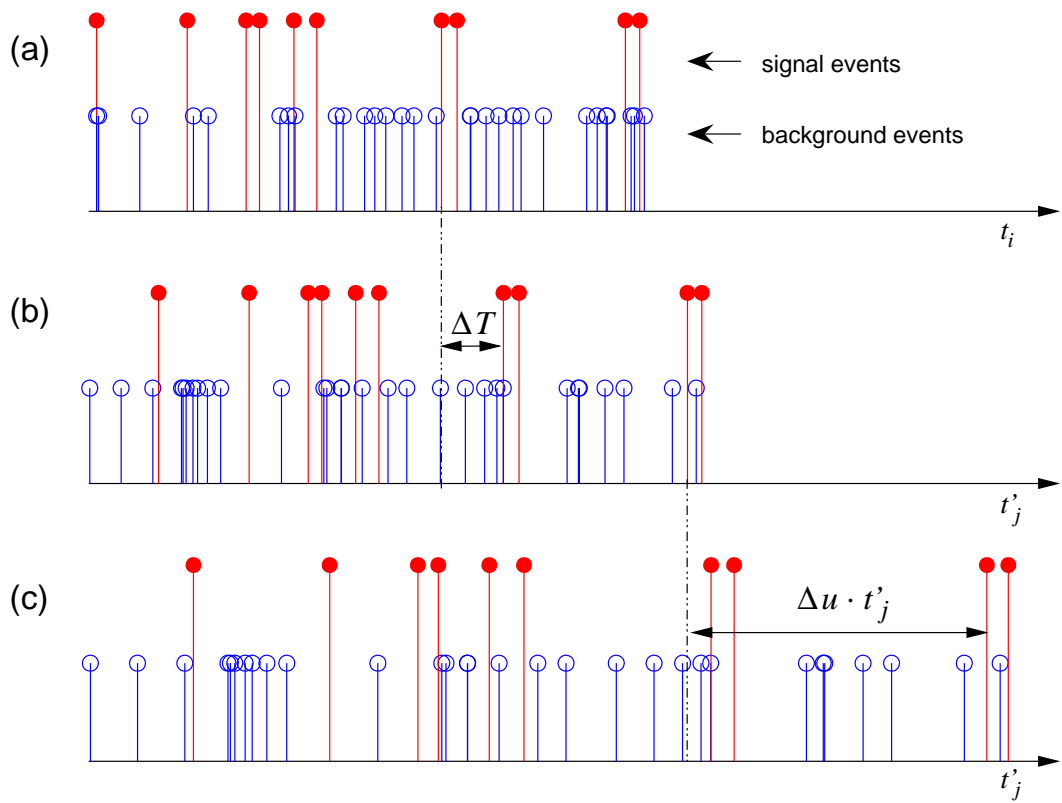}}
  \caption{\label{fig:signalexample} Effect of time offset
    and clock drift on photoevent sets. 
    Trace (a)
    represents the event set $\{t_i\}$ on side A, trace (b) an event set
    $\{t'_j\}$ on side B with a time offset $\Delta T$, but the same
    reference clock frequency. Trace (c) illustrates a set $\{t'_j\}$ with
    an additional relative frequency difference $\Delta u$ between both
    reference clocks.}
\end{figure} 

The two problems of finding time- and frequency differences from coincidence
signals in the presence of uncorrelated background events are illustrated
in figure~\ref{fig:signalexample}. Trace (a)
shows a distribution of detection events $\{t_i\}$ on side A, trace (b)
reflects the event stream on side B, assuming that there is only a time
offset $\Delta T$, but no frequency difference between the two reference
clocks. Trace (c) shows an event stream in side B both under presence of a
time offset and a frequency difference. For convenience, we describe the
relative 
frequency difference by a quantity $\Delta u$, such that the detection times
$t, t'$ on both sides due to identified photon pairs are connected via
\begin{equation}\label{eq:dispersionrelation}
t' = (t + \Delta T)\cdot (1 + \Delta u).
\end{equation}

We now estimate  how accurately $\Delta T$ and $\Delta u$ need
to be determined. In a practical QKD implementation, the two timestamping
clocks are coarsely 
synchronized with conventional means (e.g. using an NTP protocol 
\cite{mills:91}), so it can be assumed that $\Delta
T$ will not exceed a few 100\,ms. A coincidence time window may be about 1 to
5\,ns wide, fixing the uncertainty in $\Delta T$ to be small enough to start
coincidence time tracking as sketched above. Thus,
$\Delta T$ needs to be known with a precision of a few $10^{-9}$,
corresponding to an information of about 26 to 28\,bit. For the tracking
algorithm to take over, the relative frequency difference $\Delta u$ needs to
be also known to an uncertainty of $10^{-8}$ to $10^{-6}$. An upper bound for
$|\Delta u|$ can be chosen to match a typical accuracy of standard crystal
oscillators (e.g. $10^{-4}$).
Thus, $\Delta u$ of the two clocks needs to be found with a precision of
$10^{-2}$ to $10^{-4}$, equivalent to an information of 7 to 14 bits. 

\section{Finding the time offset}\label{sec:simpletimediff}
We first explain the algorithm to find the time offset $\Delta T$,
assuming the two reference clocks run at the same frequency ($\Delta
u=0$). Two streams of detection events $\{t_i\}$ and $\{t'_j\}$ on both sides
are translated into detection time functions 
\begin{equation}
  a(t)=\sum\limits_i\delta(t-t_i)\,,\quad
  b(t)=\sum\limits_j\delta(t-t'_j)\,.
\end{equation}
The cross correlation between these two functions,
\begin{equation}
c(\tau)=(a\star b)(\tau):=\int a(t)b(t+\tau)\,dt\,,
\end{equation}
has a peak at $\tau=\Delta T$ due to the correlated
photodetection events on top of an unstructured but noisy base line from
independent background detection events at both sides. The time offset
$\Delta T$ 
is thus simply found by searching for the maximum in $c(\tau)$.
In practice, $c(\tau)$ is efficiently obtained from the timing sets via fast
Fourier transformations (FFT) and their inverse,
\begin{equation}\label{eq:crosscorr}
  c(\tau)={\cal F}^{-1}\left[{\cal F}^*\left[a\right]\cdot{\cal F}\left[b\right]\right]\,,
\end{equation}
with discrete arrays for $a$, $b$ and $c$ of length $N$ (typically a power of
2). The high resolution
necessary for $\Delta T$ (28 bits) renders a direct calculation
impractical.
It is possible, however, to obtain the coarse and fine
part of $\Delta T$ separately with much smaller sample sizes. To illustrate
how this works, we take the timing events $\{t_i\}$, $\{t'_j\}$ captured
during an acquisition time $T_a$, and map them onto the discrete arrays
$\{a_k\},\{b_k\}$ with a time resolution $\delta t$:
\begin{equation}\label{eq:discreteset}
  a_k = \sum\limits_i\delta_{k,\lfloor (t_i/\delta t) \,\rm{mod}\, N\rfloor}\,,\quad
k=0,\ldots N-1\,,
\end{equation}
and $\{b_k\}$ accordingly. This is an efficient process which requires
visiting each entry $t_i$ only once.
The cross correlation array $\{c_k\}$ is obtained by the
discrete version of equation~(\ref{eq:crosscorr}), and its maximum
located by a subsequent linear search in $\{c_k\}$. 
If the cross correlation peak can be identified correctly, the result
$k_{max}$ reflects $\Delta T$ up to a resolution $\delta t$, and 
modulo $N\delta t$. Thus, applying this method with two different
resolutions $\delta t$ leads to a final $\Delta T$ with a resolution of 26
to 28 bit, while the individual FFTs are carried out at a moderate size of 
$N=2^{19}$ or less. The complete code for this procedure is available as open
source \cite{googlecode}.

It is beneficial to consider the influence of uncorrelated background events
in this peak finding 
process. We assume a signal rate $r_s$ of true coincidences, and
background rates $r_1$ and $r_2$ on both sides. The discrete arrays
$\{a_k\},\{b_k\}$ are built up from timestamps $\{t_i\},\{t'_j\}$ in a 
collection interval $T_a$. The cross correlation peak  will be
made up by $r_sT_a$ event pairs at the index $k_{max}$, while the $r_1r_2T_a^2$
background event pairs are homogeneously distributed over all $N$ entries in
 $\{c_k\}$ following a Poisson distribution. The peak can be
identified with sufficient confidence if its statistical significance $S$, here
defined as the ratio between the peak height above the base line and the
standard deviation of the latter,
\begin{equation}\label{eq:statsigni}
S(k):={{c_k-\overline{c_k}}\over\sqrt{\,\overline{(c_k-\overline{c_k})^2}}}\,,
\end{equation}
exceeds a certain numerical value. With the above rates, the peak value
arising from signal pairs is 
\begin{equation}\label{eq:statsigni2}
S_{p}={r_sT_a\over\sqrt{r_1r_2T_a^2/N}}=\sqrt{r_s^2N\over r_1r_2}\,.
\end{equation}
 If we approximate the fluctuations on the base line of $\{c_k\}$ by a Gaussian
 distribution, the probability $\epsilon$ that a base line fluctuation
 gives rise to the largest value $S_{max}$, and thus leads to a wrongly
 identified location of the cross correlation peak, is given by 
\begin{equation} \label{eq:confidence}
\epsilon = \mathcal{P} \left( S_{max}>S_{p} \right) \approx {N\over2}\left(1-\,\textrm{erf}\,{S_p\over\sqrt{2}}\right)\,.
\end{equation}
A numerical evaluation of this quantity (see table~\ref{tab:confidence})
shows that for $N<10^7$, $S_p>6$ leaves less than 1\% probability of
misidentifying the peak. Since $S_{max}$ can be
directly estimated out of  $\{c_k\}$, it forms a good basis to gauge the
success of the peak finding procedure in practice.

\begin{table}
\caption{\label{tab:confidence}
Connection between the probability $\epsilon$ of wrong peak
  identification, bin number $N$ and statistical significance $S$ of a peak. }
\vspace*{2mm}
\centerline{\begin{tabular}{c||l|l|l|l|l|l|l|l|l}
$\epsilon/N$&$10^{-4}$&$10^{-5}$&$10^{-6}$&$10^{-7}$&$10^{-8}$&$10^{-9}$&$10^{-10}$&$10^{-11}$&$10^{-12}$\\
\hline
$S$&3.72&4.26&4.75&5.12&5.61&6.00&6.36&6.71&7.03
\end{tabular}
}
\end{table}

Care should be taken that events acquired over a time $T_a$ are
uniformly distributed over the interval $N\delta t$ in the binning procedure of
equation~(\ref{eq:discreteset}). Specifically,  $T_a/(N\delta t)$ should be an
integer number. Otherwise, uncorrelated background events are subject to
an effective envelope and do not lead to a flat base line in the cross
correlation array, so determination of $\overline{c_k}$ and subsequent peak
finding becomes difficult. This problem can also be addressed by removing the
lowest Fourier components in equation~(\ref{eq:crosscorr}) before the back
transformation.

From equation~(\ref{eq:statsigni2}) it can be seen that for given rates
$r_1$, $r_2$, and $r_s$, the only way to increase the success probability is to
increase the number of time bins $N$. For $r_1=r_2=$100\,kcps, $r_s$=1\,kcps,
we need $N\approx360\,000<2^{19}$ to exceed $S=6$.
Furthermore, the frequency difference of the reference
clocks needs to be very small. A time stretch $T_a\Delta u$ between the two
clocks over an acquisition time $T_a$ exceeding the targeted resolution
$\delta t$  for $\Delta T$ reduces the statistical significance of the
coincidence peak below a useful level. With parameters $\delta t$=2\,ns
and $T_a$ of a few seconds for the experiments carried out in
\cite{marcikic:06, ling:08}, reference clocks with a relative frequency
difference $\Delta u<10^{-9}$ were necessary, which were provided in 
the form of Rubidium oscillators. 

\begin{figure}
\begin{center}
\includegraphics[scale=1]{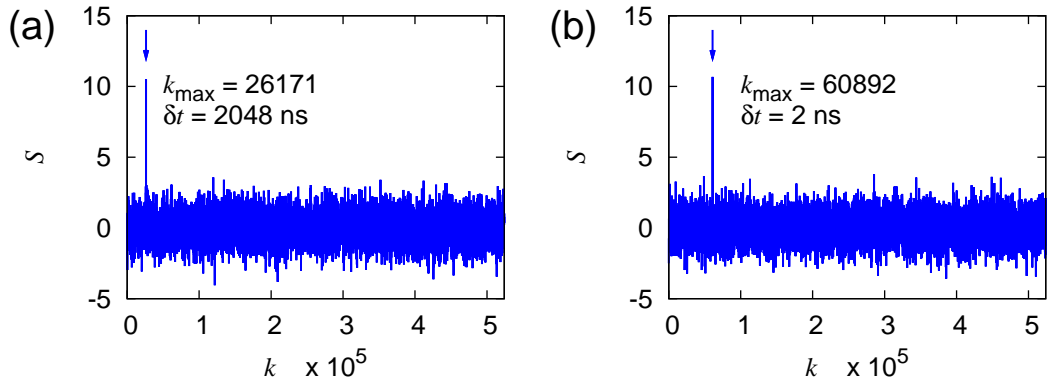}
\caption{\label{fig:convolutionpeak} Cross correlation arrays $\{c_k\}$ of
    photo events acquired over $T_a\approx1.05$\,s, normalized to a statistical
    significance $S$ as defined in equation~(\ref{eq:statsigni}) with
    $N=2^{19}$. The $k_{max}$ for two time resolutions $\delta t$ in (a) and
    (b) lead to a value $\Delta T=53\,599\,160\pm2\,$ns. All traces are sampled
    down by a factor 64.} 
\end{center}
\end{figure}

Figure~\ref{fig:convolutionpeak} shows the result of typical correlation arrays
$\{c_k\}$ (rescaled in terms of $S$) from an experiment with event rates
$r_1\approx68\,$kcps, $r_2\approx56\,$kcps and $r_s\approx1280\,$cps. Here,
$N=2^{19}$ was chosen, and time resolutions $\delta t=2\,048$\,ns for the
coarse, 
and $\delta t=2$\,ns for the fine resolution. The peak exhibits $S>10$ for both
resolutions, and the resulting time offset is $\Delta T=53\,599\,160\pm2$\,ns.

\section{Finding the time offset in the presence of a frequency difference}\label{sec:twodiffs}
The only reason to use reference clocks with a relative frequency accuracy
better than $10^{-9}$ with the presented algorithms is to determine the initial
time offset $\Delta T$ with a resolution on the order of 1\,ns. Knowledge
of the frequency difference to that accuracy and a reasonable stability is
sufficient for tracking, so it is desirable to extract this information out of
the timing events efficiently.

Finding both the time and frequency difference from the recorded timing signals
$\{t_i\}, \{t'_j\}$ in the presence of uncorrelated background events is
equivalent of identifying a line in a $(t_i,t'_k)$ plane of all
possible pair events. This well-known pattern recognition problem is formally
solved by the Hough transformation \cite{hough:59}, which maps
the pair time distribution $\{(t_i,t'_k)\}$ onto the parameter space
$\{(\Delta T, \Delta u)\}$. As in the cross correlation method in the previous
section, the pair $(\Delta T, \Delta u)$ searched for is the peak coordinate
in the parameter space. However, we did not find an equally efficient
high-resolution solution as for the one-dimensional problem in
section~\ref{sec:simpletimediff}. 

A simpler method for determining $\Delta u$ is to estimate time
offsets $\Delta T_1, \Delta T_2$ during relatively short acquisition
intervals $T_a$, shifted by a time $T_s > T_a$ with the method described in
the previous section. 
The change in time offsets between these probe intervals is connected with
the relative frequency difference $\Delta u$ via
\begin{equation}\label{eq:freqdrift1}
\Delta u={\Delta T_1-\Delta T_2\over T_s} \label{eq:freqdiff}\,.
\end{equation}
However, it is necessary to reliably obtain the time offsets on the
two sampling intervals -- which itself is only possible with clocks with a
sufficiently small $\Delta u$.  We now evaluate under which conditions this
cross correlation step will succeed in finding a time offset $\Delta T$.

For two clocks with $\Delta u=0$, the contribution of correlated events will
all end up in a single time bin in the discrete correlation array $\{c_k\}$. For
$\Delta u\neq0$, the correlated events will spread out over roughly $m=\Delta
u T_a/\delta t$  bin indices $k$. This reduces not only the statistical
significance $S$ for identifying a maximum, but also increases a
timing uncertainty which in turn leads to an uncertainty in determining the
frequency difference $\Delta u$ according to equation~(\ref{eq:freqdrift1}).

In order to identify the correlation peak with sufficient confidence
$1-\epsilon$ according to equation~(\ref{eq:confidence}), the statistical
significance should exceed a threshold $S_{th}\approx6$. For this, the timing
resolution $\delta t$ may have to be increased, forcing the true coincidences
in less bins $k$, up to to $\delta t=\Delta u T_a$, or equivalently
$N=1/\Delta u$. This, together with equation~(\ref{eq:statsigni2}) for 
the statistical significance, leads to an expression for the maximally
acceptable frequency difference 
\begin{equation}\label{eq:maxdeltau}
  \Delta u_{max}= {r_s^2\over r_1 r_2 S_{th}^2}
\end{equation}
for this strategy. 
In practice, the choice of a suitable size $N$ for the correlation array
$\{c_j\}$ has not to be done before the time-consuming step of the cross
correlation in 
equation~(\ref{eq:crosscorr}). If with an initially chosen resolution $\delta t$
the peak is not found, the array $\{c_j\}$ can be either re-partitioned in
larger bins of width $\delta t'$, or equivalently exposed to a moving average
procedure until a 
statistically significant correlation peak is identified. 
Once the time offset $\Delta T$ is known with an accuracy $\delta t'$,
the relative frequency difference $\Delta u$ is known with an accuracy
\begin{equation}\label{eq:frequncertainty}
\delta u\approx\sqrt{2}{\delta t'\over T_s}=\sqrt{2}{T_a\over NT_s}\,. 
\end{equation}
In the low signal limit, where $N$ has to be chosen large enough to identify a
peak at all, this results in a worst-case accuracy of $\delta
u=\sqrt{2}T_a/T_s$.

\begin{figure}
\begin{center}
\includegraphics[scale=1]{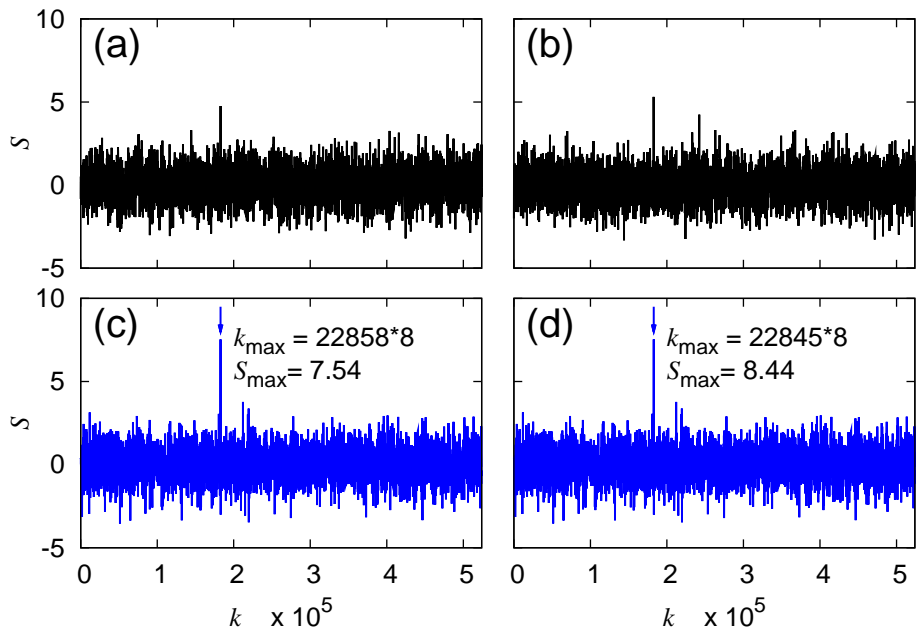}
\caption{\label{fig:realdata1}Correlation arrays for photoevents acquired
    with slightly detuned reference clocks. Traces (a) and (b) show arrays
    taken during acquisition time slots $T_a\approx268$\,ms (1\,s for events
    at side B), separated by $T_s\approx1.074$\,s and $\delta t=2.048\,\mu$s
    -- correlation peaks cannot be identified with sufficient
    significance. Traces (c) and (d) show the arrays after summing every 8
    adjacent bins, revealing a moving correlation peak. All traces are sampled
  down.} 
\end{center}
\end{figure}

We illustrate this method with experimental timing sets obtained
from a non-optimized down conversion source with moderate background event rates
($r_1\approx r_2\approx 77$\,kcps, $r_s\approx15$\,kcps). Both timestamp
units were referenced to crystal oscillators with a nominal
frequency accuracy of 100\,ppm.

With two segments of $T_a=2^{28}$\,ns $\approx 268$\,ms and a separation of
$T_s=4T_a\approx1.074$\,s, convolution arrays $\{c_k\}$ were generated with a
binning resolution of $\delta t=2.048\,\mu$s. Since there were slowly varying
changes in the background rates, the 20 lowest frequency entries (and their
mirrors) were set to
0 before back transformation to $\{c_k\}$, resulting in a smooth base line.
The results are shown in the top traces of
figure~\ref{fig:realdata1}, without any significant correlation
peaks. A subsequent re-binning with an effective width of $\delta
t'=16\,384\,$ns reduced the noise level of the background sufficiently to
allow the identification of the correlation peaks, resulting in time
offsets of $\Delta T_1=\Delta T=374\,505\pm16\,\mu$s, $\Delta
T_2=374\,292\pm16\,\mu$s, and subsequently in $\Delta u=\left(
  1.98\pm0.21\right) \cdot10^{-4}$.

\section{Iterative procedure to decrease timing and frequency uncertainty }\label{sec:iteration}
The simple method for obtaining both $\Delta T$ and $\Delta u$ by analyzing
correlations does typically not provide a sufficiently low uncertainty to
start the tracking algorithm described in section~\ref{sec:tracking}.
Therefore, additional steps are required. Knowledge of $\Delta u$ with
uncertainty $\delta u<\Delta u$, the linear dependency of the timing
uncertainty $\delta t$ from $\Delta u$ according to 
equation~(\ref{eq:frequncertainty}), and the time offset
$\Delta T$ with some accuracy suggest an iterative method for this
purpose:

A set $\{\tilde t'_j\}$ is prepared from $\{t'_j\}$, which is corrected with
the initial values $\Delta T$ and $\Delta u$ (obtained as in
section~\ref{sec:twodiffs}) via
\begin{equation}\label{eq:compensation}
  \tilde t'_j=(\tilde t'_j+\Delta T)\cdot(1+\Delta u)\,.
\end{equation}
With this set and the original set $\{t_i\}$, new values for $\Delta u$ and 
$\Delta T$ are obtained. The reduction in uncertainty $\delta u$ is given
by the ratio $T_s/T_a$  according to equation~(\ref{eq:frequncertainty}), and
is somewhat below an order of magnitude, or about 3 bit. This can be iterated,
finally leading to values $\Delta T$ and $\Delta u$
with the targeted uncertainties (see appendix for the
explicit algorithm).

\section{A faster algorithm for finding the fine time offset}
\label{sec:finepart}
The iterative method in jointly finding $\Delta T$ and $\Delta u$ with
sufficient accuracy converges only slowly because the time separation $T_s$
is typically not be very much larger than an acquisition time interval $T_a$
to keep the initial time for finding the coincidences low.

Once initial values for $\Delta u$ and $\Delta T$ are found, an alternative
algorithm can be used: We begin with event pairs sets $\{t_i\}$ and
$\{\tilde t_j'\}$, where the latter is corrected via
equation~(\ref{eq:compensation}) similarly as before. If the
timing resolution $\delta t$ for discretization is small enough (i.e.,
$r_{1,2}\delta t<1$), the arrays $\{a_i\}$ and $\{b_j\}$ are
only sparsely populated. For  $\tilde t'_j\in [0,T_b]$ with $T_b=\delta
t/\delta u$, signal events lead to coincidences in time bins with the
same time bin indices $k=k'$. The sparse population of the arrays $\{a_k\},
\{b_{k'}\}$ ensures then that the presence of a condition $a_k=b_{k'=k}=1$ is
very likely due to a true coincidence.
For those pair events, the instantaneous time difference $\Delta t=t_i-\tilde
t'_j$ due to an inaccurately known $\Delta T$ and $\Delta u$ can be determined
with an accuracy limited only by the time resolution of the detection system
(typically dominated by detector timing jitter).
 Analysis of instantaneous time differences $\Delta t$ over the
time interval $T_b$  finally reveals the parameters $\Delta T$ and $\Delta u$
with the intrinsic resolution of the system, after
which the tracking algorithm can take over. 

\begin{figure}
\begin{center}
\includegraphics[scale=1]{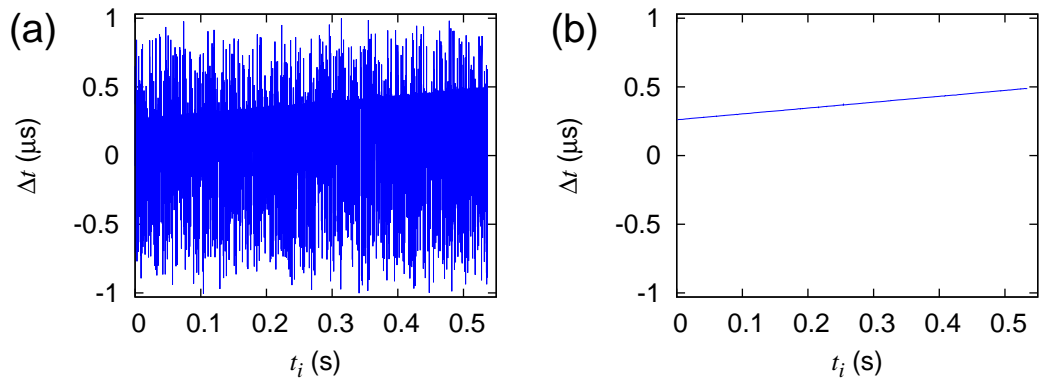}
\caption{\label{fig:sorting3a} (a) Time differences $\Delta t$ between event
  pairs on both sides falling into the same time bin after
    pre-compensation with approximate $\Delta T, \Delta u$. A large fraction
    of the pairs appear on a line, with accidental 
    coincidental pairs contributing to the noise of the figure. The
    differences fall in the range $\pm\delta t/2$, and are known with a high
    precision. (b) Dropping adjacent pairs with excessive differences leaves a 
    line which can be used to extract the final $\Delta T, \Delta u$. } 
\end{center}
\end{figure}

A distribution of time differences generated with this method from the time
stamps used in section~\ref{sec:twodiffs} is shown in
figure~\ref{fig:sorting3a}(a). 
With one more iteration of the correlation algorithm, values $\Delta
T=374\,592.8\pm1.0\,\mu$s and $\Delta u=2.0113\pm0.0014\cdot10^{-4}$ were
obtained to prepare the corrected set $\{\tilde t'_j\}$. The binning window for
identifying coincidences was chosen as $\delta t=1.024\,\mu$s. Out of the
$2^{19}$ bins for both sets $\{t_i\}$ and $\{\tilde t'_j\}$, about
20\,500 were occupied with one entry, and 75 and 191 with two events each; the
rest were empty, thus forming sparse arrays.

One can visually identify a line structure, starting at about $+0.25\delta t$,
and increasing to about $+0.5\delta t$
towards the last of the 7839 coincidence candidates. Several of them are
located away from this line, corresponding to bin pairs with
accidental coincidences. The instantaneous time difference $\Delta t$ of true
coincidences increases 
only slowly with the binning index $k$, whereas the accidental coincidences
can take arbitrary values. Thus, adjacent coincidence pairs with
bin indices $k<k'$ with a difference $\Delta t_{j(k')}-\Delta
t_{i(k)}$ in their instantaneous time difference exceeding a modulus of
$(t_{j(k')}-t_{i(k)})/N$ are likely
to contain at least one accidental coincidence. In a cleaning step, such
pairs of adjacent candidates are simply removed. This step left only
a small number of 384 coincidence candidates in the list, apparently without
any accidentals (see figure~\ref{fig:sorting3a}(b)). A linear fit with a
model $\Delta t=\Delta T'+\Delta u' t_i$ with data from the remaining pairs
returns offset correction parameters $\Delta T'=260.5\pm0.05$\,ns and $\Delta
u'=4.270\pm0.002\cdot10^{-7}$. The error intervals from the fit appear overly 
optimistic, so we reconsider them, assuming a timing uncertainty of 1\/ns in
instantaneous measurements. We finally arrive at $\Delta
T=374\,593\,062\pm1$\,ns and $\Delta u=-200\,789\pm1.4\cdot10^{-9}$.

\begin{figure}
\begin{center}
\includegraphics[scale=1]{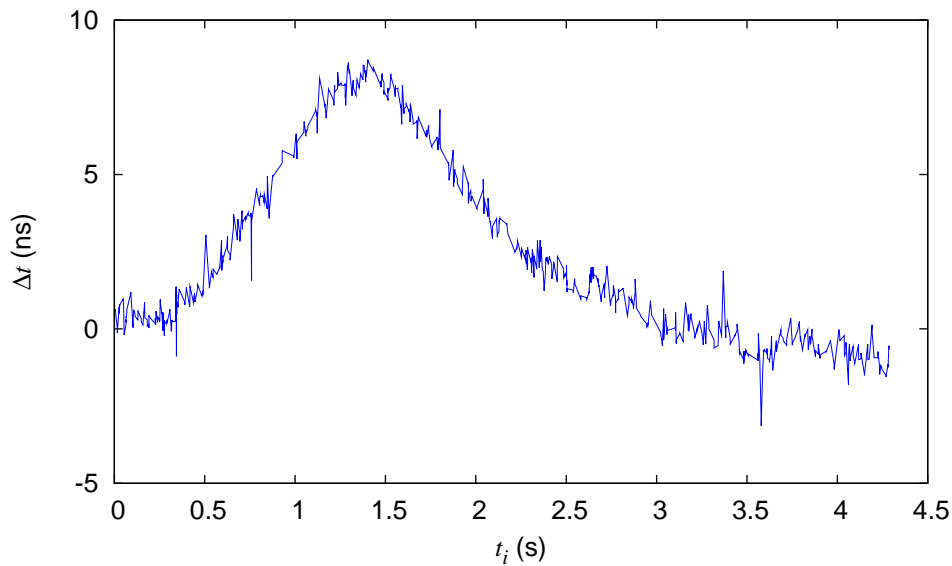}
\end{center}
\caption{\label{fig:tracking} Time differences for coincidences after
    correction of the event times at side B with $\Delta u, \Delta T$. The
    variation can now be followed by a coincidence tracking scheme described
    in section~\ref{sec:tracking}. } 
\end{figure}
 
These values are sufficient to start the tracking algorithm
sketched in section~\ref{sec:tracking}. Figure~\ref{fig:tracking} shows the
evolution of the instantaneous time difference $\Delta t$, derived in a
similar way as figure~\ref{fig:sorting3a}(b), but with the data corrected by the previously obtained constants
$\Delta T, \Delta u$ for modeling the reference clock difference. One can
recognize the slow drift of the reference oscillators,
suggesting a stability around $10^{-8}$ on a time scale of a second.

In conclusion, we presented an algorithm to remotely identify correlated
photon pairs generated in a SPDC process from a stream of detection times
without the need for a dedicated hardware channel, very stable and accurate
pair of reference clocks or a central clock source, which may expose the
classical infrastructure of a QKD system to a risk of compromise.
This greatly reduces the technical complexity of entanglement-based QKD
systems, making this part of an effort to simplify hardware by using
intrinsic information in the photon pairs and bringing it closer to
applications.

This work is supported by the National Research Foundation \& Ministry of
Education, Singapore, and partly by a joint program of quantum information
research between DSO and NUS.

\section*{Appendix: Iterative algorithm for finding time- and frequency difference}
\label{sec:appendix}
An explicit algorithm for obtaining time- and frequency differences $\Delta T$
an $\Delta u$  from time sets $\{t_i\}, \{t'_j\}$ with high precision
comprises the following steps:
\begin{enumerate}
\item Choose the limits for the maximum expected frequency and time differences $\Delta u_{max}$ and
  $\Delta T_{max}$; the observed rates $r_1,r_2$ and an expected rate $r_s$
  can be used to check with equation (\ref{eq:maxdeltau}) if this algorithm
  can be expected to  be successful.
\item Choose an acquisition time interval $T_a$ and a separation
  time interval $T_s$, e.g. $T_s=10 T_a$. 
\item Choose the smallest discretization time $\delta t$ that is compatible
  with a high chance of successfully identifying the correct peak in the cross
  correlation. Pair this with a suitable value of $N$ for generating the
  arrays $\{a_j\}$ and $\{b_k\}$ according to
  equation~(\ref{eq:discreteset}).
\item Generate the cross correlation array $\{c_k\}$ via FFT as in
  section~\ref{sec:simpletimediff}.
\item Find the index $k$ of the maximal value in $\{c_k\}$ and estimate its
  statistical significance $S$ according to equation (\ref{eq:statsigni}).
\item If $S$ is below a chosen significance limit $S_{th}$, half the size
  of the array $\{c_k\}$ by adding entries pairwise, and go back to
  step 5; this doubles the effective time resolution $\delta t'$. Otherwise,
  continue.
\item Determine $\Delta T_1$ from the peak position $k_{max}$ and the effective
  time resolution $\delta t'$ of the last iteration of the previous step.
\item With the time resolution $\delta t'$ in step 6, generate discrete
  arrays $\{a_j\}, \{b_k\}$ and $\{c_k\}$ for the second sampling interval, and
  determine from there $\Delta T_2$.
\item Determine $\Delta u$ from $\Delta T_1, \Delta T_2$, and $T_s$ from
  equation~(\ref{eq:freqdiff}).
\item If $\delta t'$ in the last iteration is small enough to start the
  tracking as described in section~\ref{sec:tracking}, the algorithm is
  finished.
\item Generate a modified set of event times $\{\tilde t'_j\}$ according to
    $\tilde t' = (t'+\Delta T_1)\cdot(1+\Delta u)$
\item Choose the same $N$ as in the last FFT, but reduce the time interval
  $\delta t$ by less than the expected gain in accuracy given by
  $T_s/T_a/\sqrt{2}$; typically, this reduction factor would be 4 or 8
  corresponding to 2 or 3 bits in accuracy gain.
\item Generate new $\{a_i\},\{b_j\}$ and $\{c_k\}$ in the usual way from the
  original set $\{t_i\}$ and the modified set $\{\tilde t'_j\}$; from the peak
    position in the new $\{c_k\}$, determine the correction to $\Delta
    T_1$. Usually, this adds 2 or 3 bits in accuracy to $\Delta T_1$; proceed
    similarly for the correction to $\Delta T_2$.
\item Continue with step 10.
\end{enumerate}

\end{document}